\documentclass[aps,pra,showkeys,twocolumn,superscriptaddress]{revtex4-2}
\usepackage[utf8]{inputenc}
\usepackage{amsmath,graphicx,array,amsthm}

\usepackage[colorlinks={true}]{hyperref}
\hypersetup{colorlinks=true,linkcolor=red,citecolor=blue,urlcolor=blue}
\usepackage{subfigure}
\usepackage{indentfirst}
\usepackage{braket}
\usepackage{diagbox}
\usepackage{amssymb}
\usepackage{bbold}
\usepackage{colortbl}
\usepackage{multirow}
\usepackage{orcidlink}
\usepackage{pstricks}

\begin{document}

\title{Quantum dense coding with gravitational cat states}

\author{Saeed Haddadi \orcidlink{0000-0002-1596-0763}}\email{haddadi@semnan.ac.ir}
\affiliation{Faculty of Physics, Semnan University, P.O. Box 35195-363, Semnan, Iran}

\author{Mehrdad Ghominejad \orcidlink{0000-0002-0136-7838}}\email{mghominejad@semnan.ac.ir}
\affiliation{Faculty of Physics, Semnan University, P.O. Box 35195-363, Semnan, Iran}

\author{Artur Czerwinski \orcidlink{0000-0003-0625-8339}}
\email{a.czerwinski@cent.uw.edu.pl}
\affiliation{Centre for Quantum Optical Technologies, Centre of New Technologies, University of Warsaw, Banacha 2c, 02-097 Warszawa, Poland}

\date{\today}

\begin{abstract}
A protocol of quantum dense coding with gravitational cat states is proposed.
We explore the effects of temperature and system parameters on the dense coding capacity and provide an efficient strategy to preserve the quantum advantage of dense coding for these states. Our results might open new opportunities for secure communication and possibly insights into the fundamental nature of gravity in the context of quantum information processing.
\end{abstract}
\keywords{Quantum dense coding; gravitational cat states; quantum advantage}

\maketitle
\vspace{-1.25cm}

\section{Introduction}
Quantum communication protocols use a set of technologies and algorithms to send and receive information using quantum properties such as entanglement and quantum measurements \cite{Gisin2007,Cariolaro2015}. These protocols are used for secure communication and privacy in various fields including cryptography, telecommunications, and quantum networks \cite{Renner2008,Scarani2009,Portmann2022}. Among the several communication protocols, quantum dense coding is an archetype that uses entanglement as a resource. This concept was originally introduced by Bennett and Wiesner in the 1990s \cite{Bennett1992}.
Dense coding is one of the important concepts in quantum information theory and refers to the possibility of transmitting quantum information with greater efficiency than the transmission of single bits of information. The basic idea of dense coding is that by using entanglement between sender and receiver, the sender can send two classical bits of information with one qubit \cite{Bareno1995,Mattle1996}.

The four steps involved in dense coding protocol are as follows. In the first step, the sender Bob and the receiver Alice share a two-qubit entangled state between themselves. Then, Bob decides which of the entangled qubits to manipulate to encode the two classical bits of information. This transformation is typically performed using an appropriate quantum operation and depends on the principles of quantum superposition and entanglement. Next, with the selected qubit now in a modified state, Bob sends it to Alice. This qubit has been prepared to carry the encoded information. After Alice receives the qubit sent by Bob, they finally perform measurements on the received qubit and the other entangled qubit to extract the encoded information. By making some measurements based on the received qubit and the entangled state, Alice can determine the two classical bits sent by Bob.

In literature, several studies have both theoretically proposed and experimentally demonstrated the concept of dense coding (see, e.g., Refs. \cite{Bose2000,Hiroshima2001,Lee2002,Horodecki2012,mirmasoudi2018,yqli2019,Meher2020,Haddadi2020,Rabbou2022} and references quoted therein). Interestingly, different methods have been presented to protect the quantum advantages of dense coding under decoherence for a relatively long time \cite{Shadman2010,Quek2010,Zhang2016,Zhao2019,Haddadi2022,Huml2022}, such as choosing the proper initial channel state  \cite{Yeo2006,Yeo2008}, employing different configurations for the external noise \cite{Sun2022,Haddadi2023}, and applying quantum weak measurement (QWM) operation \cite{Tian2018,Li2021,paper20231}.

The dense coding capacity (denoted by $\chi$) is a crucial metric for assessing the effectiveness of the selected quantum communication channel. Specifically, $\chi$ quantifies the potential transmission of classical information when Bob sends a single qubit to Alice. The achievable amount of information transmission is bounded by the Holevo quantity \cite{Holevo1973}. Given that the Holevo quantity is asymptotically reachable \cite{Holevo1998}, the dense coding capacity can be defined as \cite{Schumacher1997}
\begin{equation}\label{eq1}
\chi\left(\rho_{A B}\right)=S\left(\bar{\rho}_{A B}\right)-S\left(\rho_{A B}\right).
\end{equation}

Here, $\bar{\rho}_{A B}$ is the signal ensemble average state, which is calculated as $\frac{1}{4} \sum_{i=0}^{3} (\sigma^i \otimes \mathbf{I})\rho_{AB}(\sigma^i \otimes \mathbf{I})$ where $\sigma^0=\mathbf{I}$ is identity operator and $\sigma^{1,2,3}$ are the Pauli operators. The function $S(\rho)=-\textmd{tr}(\rho \log_2 \rho)$ denotes the von Neumann entropy of the density operator $\rho$. It is important to note that the valid dense coding is achieved when $\chi\left(\rho_{A B}\right)$ is greater than one. Besides, $\chi\left(\rho_{A B}\right)_{\max }=2$ represents the optimal dense coding \cite{Hiroshima2001}.

In the context of a quantum interpretation of matter, it is possible to consider a stationary heavy particle as being in a state known as the Schr\"{o}dinger cat state or as a superposition of two physically distinct positions \cite{zhengprl2023}. These states, referred to as gravitating entities or gravitational cats (gravcats), exhibit behavior that is of fundamental importance to understand \cite{Anastopoulos2015}. Specifically, their behavior is of interest in the fields of gravitational non-local physics and macroscopic non-local events. Therefore, gravcats can be utilized to evaluate the resilience of quantum systems for practical applications in quantum information deployment. Despite the abundance of quantum gravity theories that claim to provide explanations within the framework of current physics \cite{Abdo2009,Hossenfelder2013,Modesto2017}, including those applicable to weak-gravitational and non-relativistic regimes \cite{
Anastopoulos2020,Anastopoulos2021}, there is a general lack of sufficient experimental evidence and recommendations concerning the interaction between gravitation and quantum matter. Hence, a comprehensive assessment of gravcats as a resource is needed in order to understand their behavior within quantum systems and their relevance to the field of quantum information science.

In this paper,  we put forward a simple question that we believe has not been addressed in the literature on this particular issue: how is the dense coding with gravcat states under a thermal bath? Although the influence of temperature on the quantum correlations of gravcat states in closed and open systems has been studied recently \cite{Rojas2023,Atta2023}, here we aim to examine the effect of various parameters on the quantum advantages of dense coding with these states and to provide an efficient strategy for reducing the destructive effects.

\section{Hamiltonian and Gibbs density operator}\label{sec2}
\begin{figure}[t]
  \centering
  \includegraphics[width=3in]{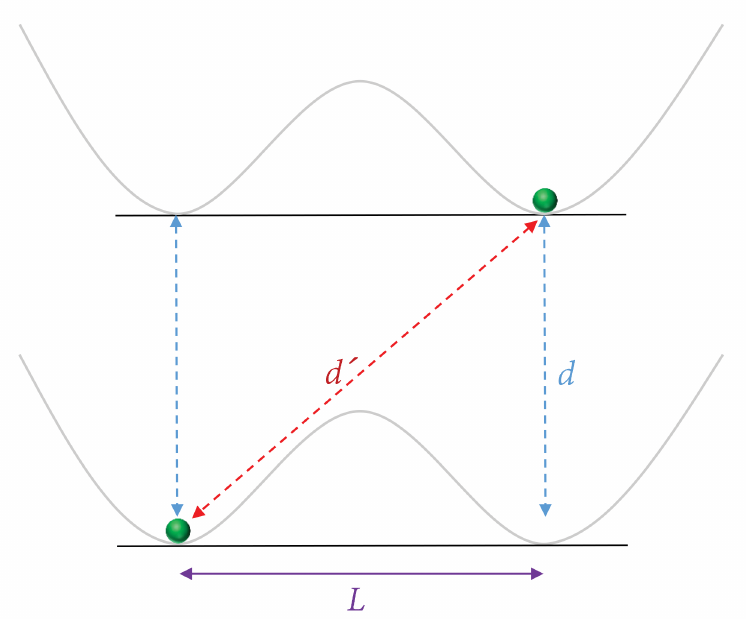}
  \caption{Geometry of the gravcats model. Each symmetric double-well potential is located along a distinct axis, while the two axes are parallel at a distance $d =\sqrt{{d^{\prime}}^2-L^2}$.}\label{figure1}
\end{figure}

Let us consider a physical model for the interaction of two gravcats, each of them corresponding to a qubit. The Hamiltonian describing this model is presented as follows \cite{Anastopoulos2020}
\begin{equation}\label{eq2}
H=\frac{\omega}{2}(\mathbf{I} \otimes \sigma^z + \sigma^z \otimes \mathbf{I})-\gamma (\sigma^x \otimes \sigma^x),
\end{equation}
where $\omega$ denotes the energy difference between the ground state and the first excited state. Moreover, $\gamma=\frac{G m^ 2}{2} \left(\frac{1}{d} - \frac{1}{d^{\prime}}\right)$ regulates the gravitational interaction coupling strength between the states with masses $m$ in which $G$ is a universal gravitational constant and $d$ ($d^{\prime}$) is the distance between the two masses when each of them is at the same (different) relative minimum (see Fig. \ref{figure1}).

The Gibbs density operator can be written as
\begin{equation}\label{eq3}
\rho=\frac{1}{Z}\exp(-H/k_{B}T),
\end{equation}
where $Z=\textmd{tr}[\exp(-H/k_{B}T)]$ is the partition function, $k_{B}$ is the Boltzmann constant ($k_{B}=1$ is considered throughout this paper), and $T$ is the absolute temperature. The operator \eqref{eq3} satisfies the following conditions: $\textmd{tr}\rho=1$, $\rho\geq0$, and $\rho=\rho^{\dag}$.

Due to the functional equation \eqref{eq3} and our Hamiltonian \eqref{eq2}, the Gibbs density matrix takes the following form
\begin{equation}\label{eq4}
\rho_T=\left(\begin{array}{cccc}
\alpha^- & \cdot & \cdot & \kappa\\
\cdot & \beta & \eta & \cdot\\
\cdot & \eta & \beta & \cdot\\
\kappa & \cdot & \cdot & \alpha^+
\end{array}\right),
\end{equation}
where the dots were used instead of zero entries and the nonzero entries are
\begin{align}\label{eq5}
\alpha^\pm= & \frac{\Theta\cosh \left(\Theta/T\right)\pm\omega  \sinh \left(\Theta/T\right)}{Z\Theta}, \quad \beta=\frac{\cosh \left(\gamma /T\right)}{Z},\nonumber \\
\kappa=& \frac{\gamma  \sinh \left(\Theta/T\right)}{Z \Theta},\quad \eta=\frac{\sinh \left(\gamma/ T\right)}{Z},
\end{align}
with $Z=2 \left[\cosh \left(\Theta/T\right)+\cosh \left(\gamma /T\right)\right]$ and $\Theta^{2}=\omega ^2+\gamma ^2$.

\section{Dense coding capacity in gravcats}\label{sec3}
In the case of our thermal state \eqref{eq4}, the average
state of the signal ensemble $\bar{\rho}_{T}$ is given by
\begin{equation}\label{eq6}
\bar{\rho}_{T}=\frac{1}{2}\left(\begin{array}{cccc}
\alpha^{-} +\beta  & \cdot & \cdot & \cdot\\
\cdot & \alpha^{+}+\beta & \cdot & \cdot\\
\cdot & \cdot & \alpha^{-} +\beta & \cdot\\
\cdot & \cdot & \cdot & \alpha^{+}+\beta
\end{array}\right),
\end{equation}
thereby, the analytical form of the dense coding capacity \eqref{eq1} is obtained as
\begin{align}\label{eq7}
\chi(\rho_{T})= & \sum_{i=+,-} a^{i}\log _2 a^{i}+\sum_{j=+,-} b^{j}\log _2 b^{j}\nonumber\\
&-\sum_{k=+,-}(\alpha^{k}+\beta)\log _2\frac{\alpha^{k}+\beta}{2},
\end{align}
where $a^{\pm}=\frac{1}{2} \left[(\alpha^{-}+\alpha^{+})\pm\sqrt{(\alpha^{-}-\alpha^{+})^2 + 4|\kappa|^2}\right]$
and
$b^{\pm}=\beta\pm|\eta|$.


Now, we turn to the investigation of the dense coding capacity for arbitrary values of $\omega$, $\gamma$, and $T$.

\section{Results and discussion}\label{sec4}

In our analysis, we consider the dense coding capacity \eqref{eq7} as a two-variable function by allowing two parameters to change in a continuous manner. As a result, we can investigate $\chi(\rho_{T})$ in three different domains, i.e., $(\gamma, \omega)$, $(T, \omega)$, and $(T, \gamma)$. These three perspectives on the dense coding capacity are presented respectively.

\begin{figure*}[hbt!]
\centering
{\includegraphics[width=80mm]{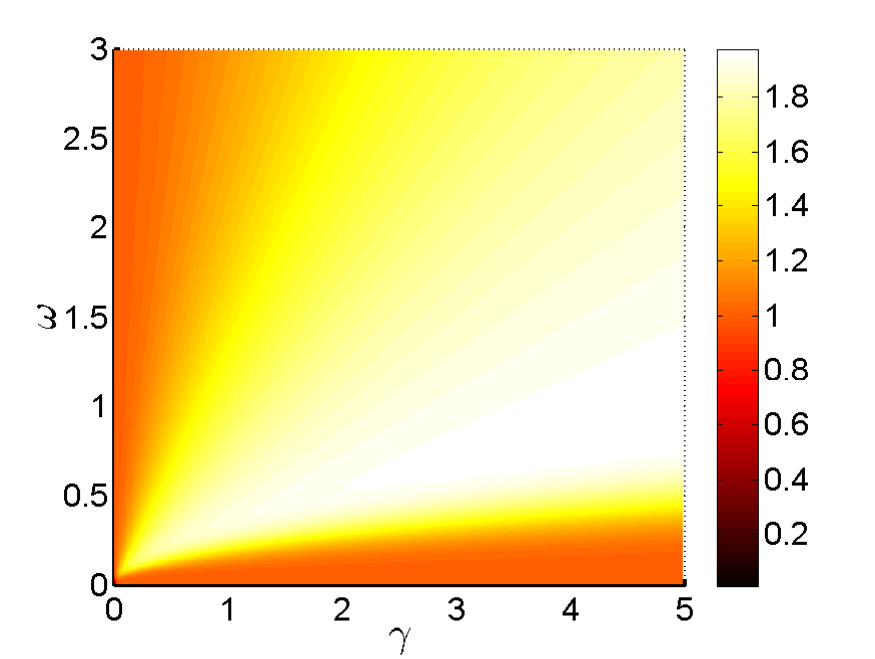}}
\put(-185,162){($ a $)}\
{\includegraphics[width=80mm]{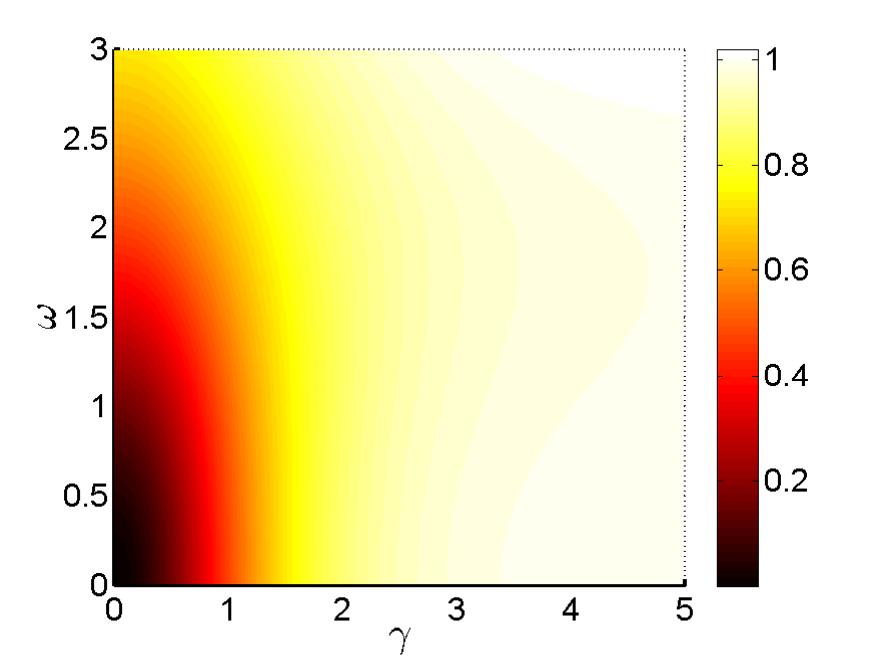}}
\put(-185,162){($ b $)}\
\caption{Dense coding capacity \eqref{eq7} in two gravcats versus $\omega$ and $\gamma$ with ($a$) $T=0.01$ and ($b$) $T=1$.} \label{figure2}
\end{figure*}

\begin{figure*}[hbt!]
\centering
{\includegraphics[width=80mm]{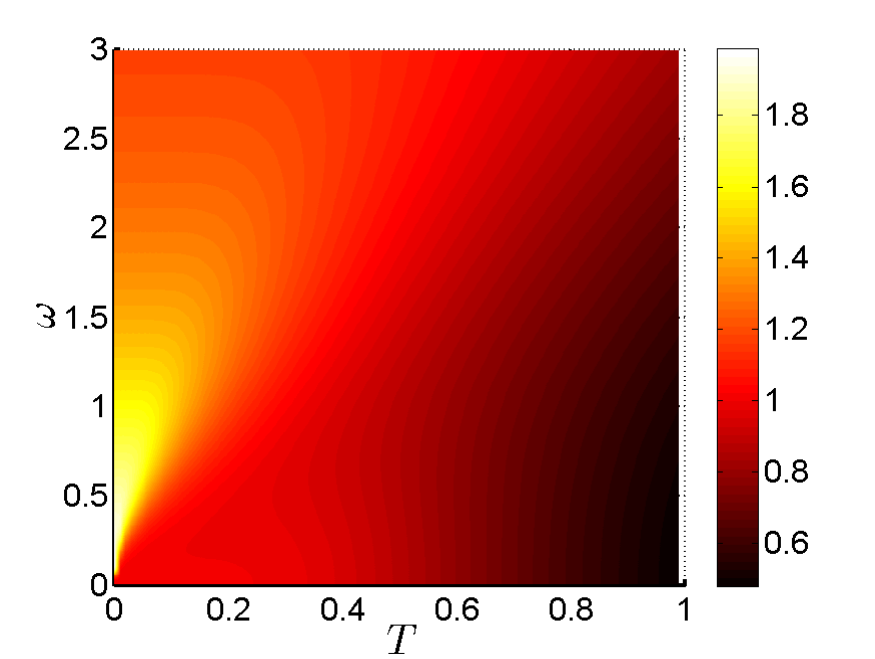}}
\put(-185,162){($ a $)}\
{\includegraphics[width=80mm]{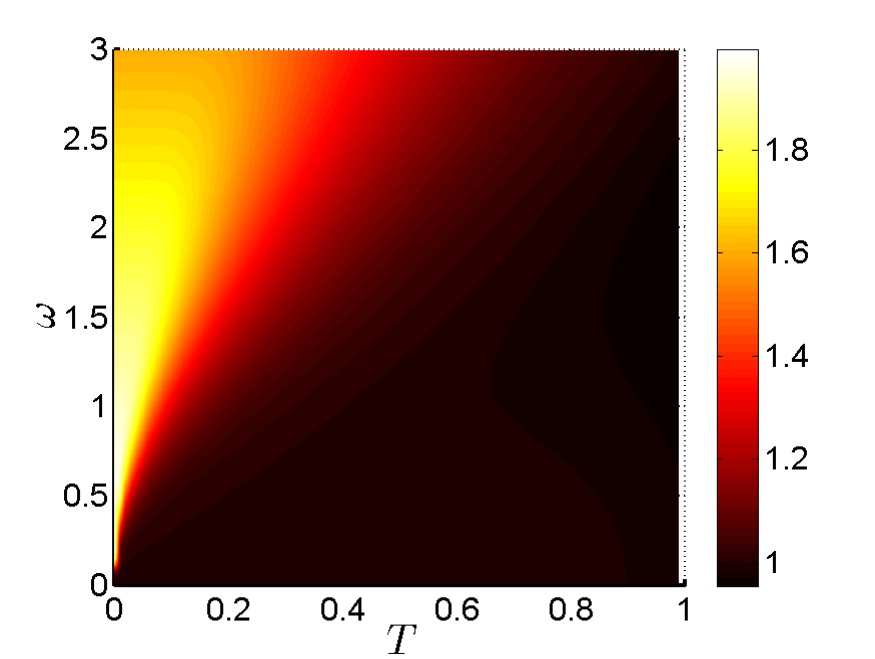}}
\put(-185,162){($ b $)}\
\caption{Dense coding capacity \eqref{eq7} in two gravcats versus $\omega$ and $T$ with ($a$) $\gamma=1$ and ($b$) $\gamma=3$.} \label{figure3}
\end{figure*}

\begin{figure*}[hbt!]
\centering
{\includegraphics[width=80mm]{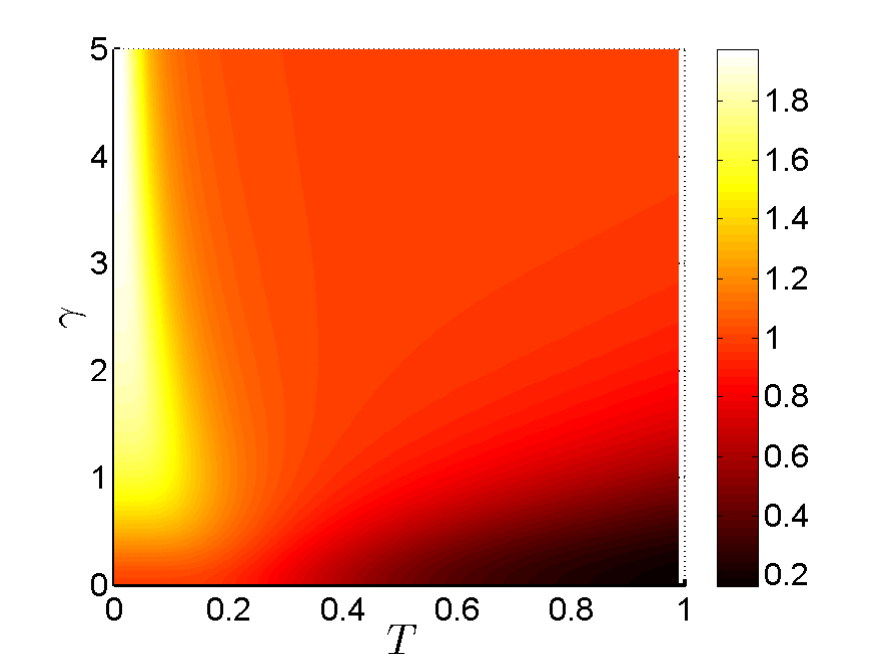}}
\put(-185,162){($ a $)}\
{\includegraphics[width=80mm]{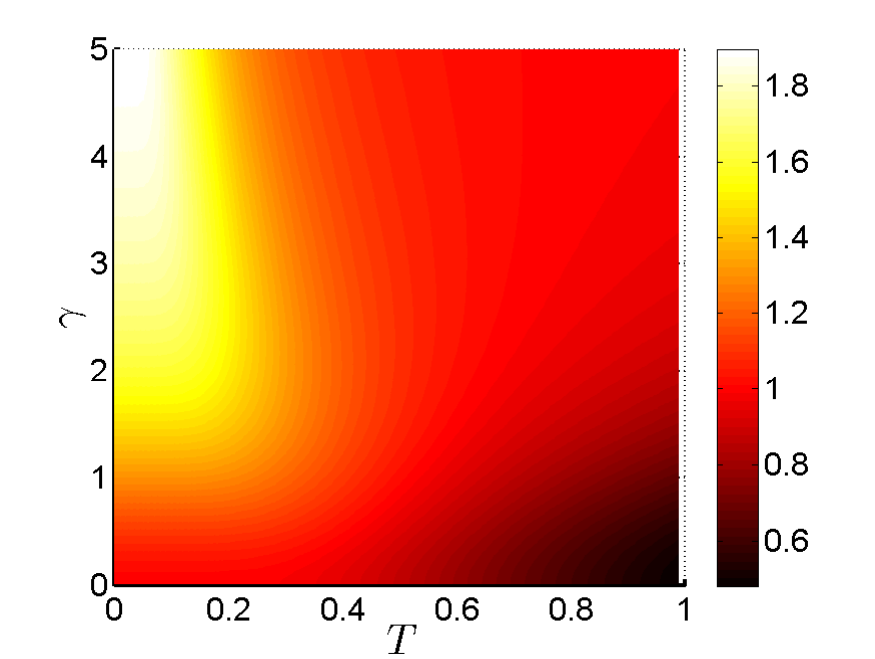}}
\put(-185,162){($ b $)}\
\caption{Dense coding capacity \eqref{eq7} in two gravcats versus $\gamma$ and $T$ with ($a$) $\omega=1$ and ($b$) $\omega=2$.} \label{figure4}
\end{figure*}

First, in Fig.~\ref{figure2}, we provide the dense coding capacity as a function of $\gamma$ and $\omega$, comparing its properties for two temperatures, i.e. $T = 0.01$ in Fig.~\ref{figure2}~($a$) and $T = 1$ in Fig.~\ref{figure2}~($b$). For the lower temperature, the behavior of the dense coding capacity is non-monotonic. For example, by choosing a fixed $\gamma$, we see that $\chi(\rho_{T})$ first increases and then decreases by growing $\omega$.

For a higher temperature, which is in Fig.~\ref{figure2}~($b$), we observe an initial decline in the dense coding capacity for relatively small values of $\gamma$ and $\omega$. This phenomenon is due to the thermal effects causing increased noise and reduced coherence in the system. However, as we increase the gravitational interaction coupling, we see that higher values of $\gamma$ lead to stronger correlations, positively impacting the dense coding capacity. Varying the energy difference $\omega$ also causes an improvement of the dense coding capacity, but this effect is not as significant as the impact of $\gamma$.

The effects of the temperature on the dense coding capacity are clearly visible in Fig. \ref{figure3}, where we consider $\chi(\rho_{T})$ as a function of $T$ and $\omega$. As temperature increases, thermal fluctuations become more significant.  In general, higher temperatures can introduce more noise and reduce the fidelity of quantum information processing. These factors result in a decrease in the dense coding capacity for larger $T$, which is visible in  Fig. \ref{figure3} ($a$) and ($b$). In the domain of $\omega$, the dense coding capacity exhibits non-linear behavior. Higher energy difference initially leads to increased capacity, as there is more energy available for quantum information processing. However, at extremely high values of $\omega$, other factors, like increased thermal effects, counteract this trend. By comparing  Fig. \ref{figure3} ($a$) and ($b$), we notice an interplay between gravitational effects and the quantum information processing capabilities. On the one hand, for $\gamma = 3$, the maximum of $\chi(\rho_{T})$ is represented by a larger and brighter area in the domain of $(T, \omega)$, but on the other hand, the higher $\gamma$ suppresses the dense coding capacity outside the maximum range.

Previously described properties of the dense coding capacity can be confirmed by analyzing Fig.~\ref{figure4}. Again, as temperature increases, thermal effects tend to dominate, leading to increased entropy in the system and causing the dense coding capacity to decrease. By comparing Fig.~\ref{figure4} ($a$) and ($b$), we see that higher energy difference $\omega$ results in more energy that is available for quantum processes, increasing the dense coding capacity when other factors remain constant. This tendency is displayed in Fig.~\ref{figure4} ($b$) by a larger region corresponding to maximum values of $\chi(\rho_{T})$. The gravitational interaction coupling strength $\gamma$ introduces a unique aspect. Strong gravitational interactions might enhance or suppress quantum effects, leading to non-linear behavior in the dense coding capacity, which is visible in Fig.~\ref{figure4}.

Our analysis demonstrates non-linear dependencies of $\chi(\rho_{T})$ on the key variables due to the complex interplay between gravitational interactions, energy levels, and thermal effects. Based on Figs. \ref{figure2}-\ref{figure4}, we see that there are regions in the parameter space where the dense coding capacity can be optimized ($\chi(\rho_{T})\approx 2)$, indicating conditions where the quantum system performs most efficiently.

\section{Dense coding capacity with weak measurement}\label{sec5}

\begin{figure*}[hbt!]
\centering
{\includegraphics[width=80mm]{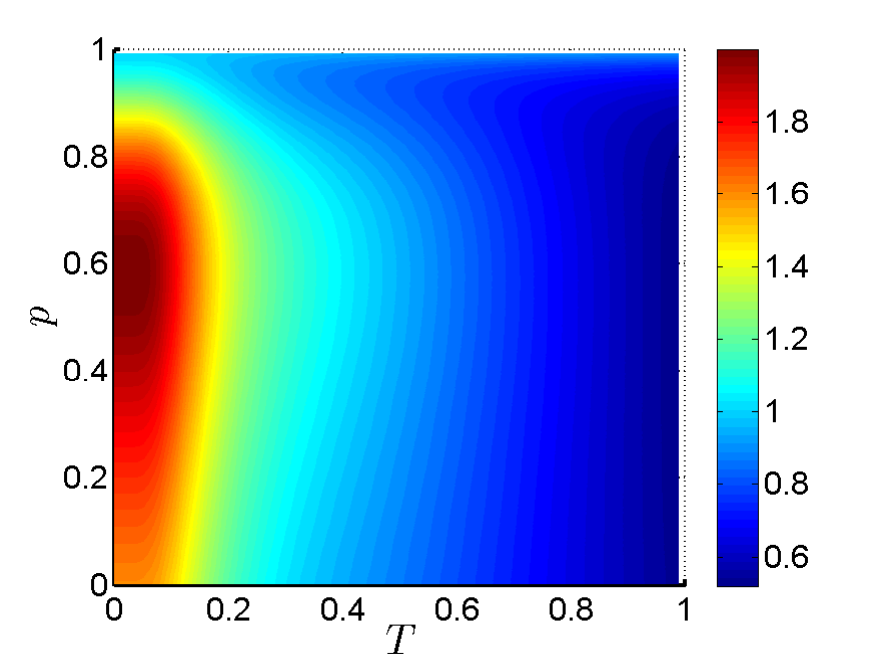}}
\put(-185,162){($ a $)}\
{\includegraphics[width=80mm]{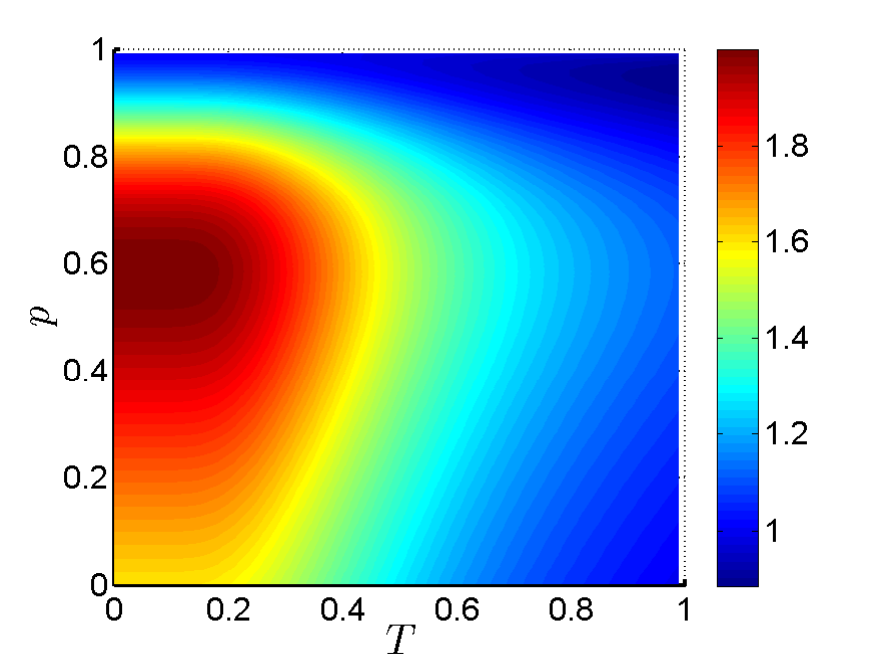}}
\put(-185,162){($ b $)}\
\caption{Dense coding capacity \eqref{dcwm} in two gravcats versus $p$ and $T$ under QWM with ($a$) $\omega=\gamma=1$ and  ($b$) $\omega=\gamma=3$.} \label{figure5}
\end{figure*}

\begin{figure*}[hbt!]
\centering
{\includegraphics[width=80mm]{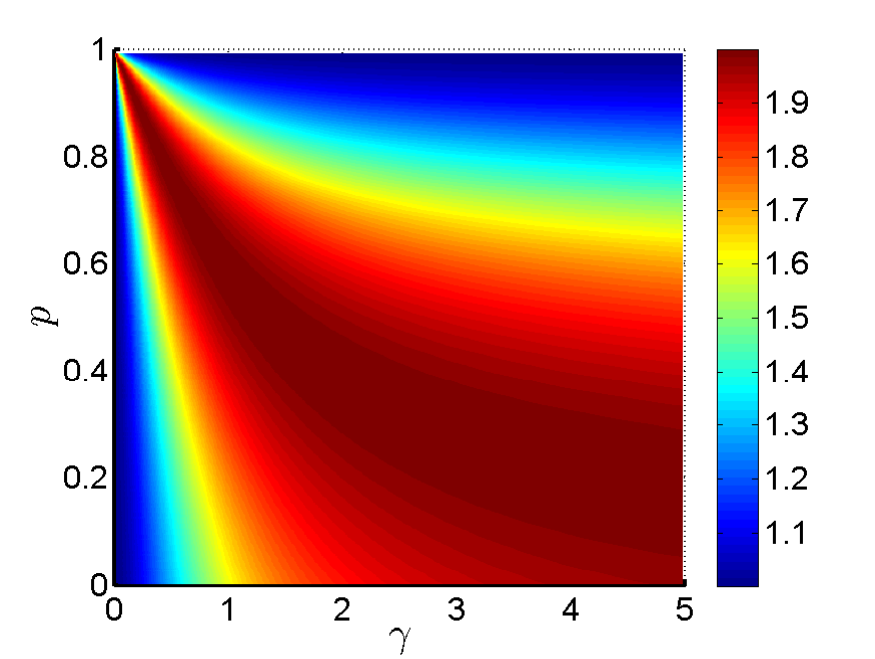}}
\put(-185,162){($ a $)}\
{\includegraphics[width=80mm]{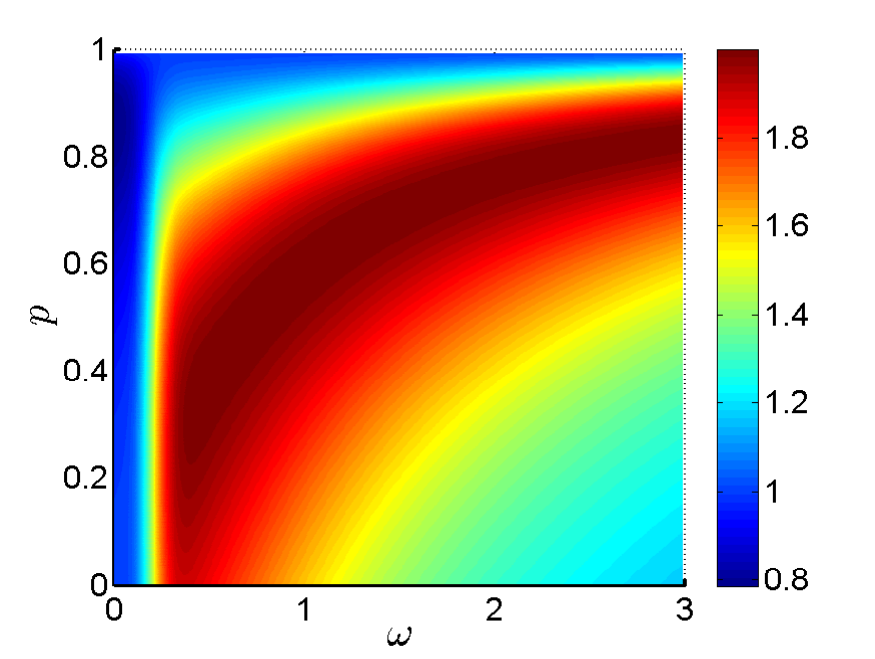}}
\put(-185,162){($ b $)}\
\caption{Dense coding capacity \eqref{dcwm} in two gravcats versus $p$, $\gamma$, and $\omega$  under QWM at $T=0.01$ for ($a$) $\omega=1$ and ($b$) $\gamma=1$.} \label{figure6}
\end{figure*}

We present here a simple practical strategy to control the capacity of dense coding using a class of uncollapsing operations, namely, QWM \cite{Aharonov1988}.
It provides a way to get information regarding a quantum system with minimal disturbance, suggesting insights into the subtleties of quantum behavior. QWM has also both theoretical significance and practical applications in precision measurements, quantum control, and quantum information processing  \cite{XiaoEPJD2013,WangPRA2014,dwang2018,Wangcpb2020,Zhangcpb2021,paper20232,paper20233}.
The QWM process can be mathematically represented by a matrix having the following structure
\begin{equation}\label{QWM}
\mathcal{Q}^{\textmd{WM}}=\left(\begin{array}{cc}
1 & \cdot \\
\cdot & \sqrt{1-p}
\end{array}\right),
\end{equation}
where $p$ denotes the measurement strength, $0\leq p \leq 1$.

Through performing the QWM on both
qubits $A$ and $B$, our thermal state \eqref{eq4} becomes
\begin{equation}\label{WMT}
\rho_T^{\textmd{WM}}=(\mathcal{Q}^{\textmd{WM}}_A \otimes \mathcal{Q}^{\textmd{WM}}_B) \rho_T(\mathcal{Q}^{\textmd{WM}}_A \otimes \mathcal{Q}^{\textmd{WM}}_B)^{\dagger }/\mathcal{P}_\textmd{s},
\end{equation}
where
$\mathcal{P}_\textmd{s}=\textmd{tr}\left[(\mathcal{Q}^{\textmd{WM}}_A \otimes \mathcal{Q}^{\textmd{WM}}_B)\rho_T(\mathcal{Q}^{\textmd{WM}}_A \otimes \mathcal{Q}^{\textmd{WM}}_B)^{\dagger }\right]$ quantifies the probability of success of the measurement.

After some calculations,
we obtain
{\small
\begin{align}\label{rowm}
\rho_T^{\textmd{WM}}&=\nonumber\\
&\frac{1}{\mathcal{P}_\textmd{s}}\left(\begin{array}{cccc}
\alpha^- & \cdot & \cdot & \kappa (1-p)\\
\\
\cdot & \beta \sqrt{(1-p)^2} & \eta \sqrt{(1-p)^2}  & \cdot\\
\\
\cdot & \eta \sqrt{(1-p)^2}  & \beta \sqrt{(1-p)^2} & \cdot\\
\\
\kappa (1-p)  & \cdot & \cdot & \alpha^+ (1-p)^2
\end{array}\right),
\end{align}}
with $\mathcal{P}_\textmd{s}=\alpha^- +2\beta\sqrt{(1-p)^2} +\alpha^+ (1-p)^2$.

By combining Eqs. \eqref{eq1} and \eqref{rowm}, the capacity of dense coding with weak measurement can be
obtained as

\begin{align}\label{dcwm}
\chi(\rho_{T}^{\textmd{WM}})= & \sum_{i=+,-} c^{i}\log _2 c^{i}+\sum_{j=+,-} d^{j}\log _2 d^{j}\nonumber\\
&-\nu\log _2\frac{\nu}{2}
-\mu\log _2\frac{\mu}{2},
\end{align}
where
\begin{align}
  2c^{\pm} \mathcal{P}_\textmd{s} =& \pm\sqrt{[\alpha^{-}-\alpha^{+}(1-p)^2]^2 + 4|\kappa(1-p)|^2}\nonumber\\
&+[\alpha^{-}+\alpha^{+}(1-p)^2]
,\nonumber
\end{align}
$$d^{\pm} \mathcal{P}_\textmd{s} =(\beta\pm|\eta|)\sqrt{(1-p)^2},$$
$$\nu \mathcal{P}_\textmd{s} =\alpha^{-}+\beta\sqrt{(1-p)^2},$$
and
$$\mu \mathcal{P}_\textmd{s}
 =\alpha^{+}(1-p)^2+\beta\sqrt{(1-p)^2}.$$

To discuss the properties of the dense coding capacity under QWM, we plot $\chi(\rho_{T}^{\textmd{WM}})$ given by \eqref{dcwm} in three domains: $(T, p)$, $(\gamma, p)$, and $(\omega, p)$. This approach allows us to investigate how the measurement strength $p$ affects the dense coding capacity.

First, in Fig.~\ref{figure5}, we have plotted $\chi(\rho_{T}^{\textmd{WM}})$ versus $T$ and $p$ for two combinations of the remaining parameters, i.e., $\omega = \gamma = 1$ [Fig. \ref{figure5} ($a$)] and  $\omega = \gamma = 3$ [Fig. \ref{figure5} ($b$)]. As for the impact of temperature, we see again that the dense coding capacity declines as we increase $T$, which is in agreement with the earlier analysis. Moreover, the effects of the measurement strength on the dense coding capacity vary with temperature, showing non-monotonic behavior. We observe that there is an optimal range of measurement strength $p$ that maximizes the dense coding capacity at different temperatures. Finally, by comparing Fig. \ref{figure5} ($a$) and (b), we conclude that the dense coding capacity becomes more robust against thermal effects for greater values of $\omega$ and $\gamma$. This means that we can protect the dense coding capacity against temperature and obtain a larger optimal range if we increase the energy difference and the gravitational interaction strength.

Next, Fig.~\ref{figure6} ($a$) shows the dense coding capacity versus $\gamma$ and $p$. Depending on the interplay between weak measurements and gravitational interactions, one can observe enhancements or suppressions of dense coding capacity. In general, when increasing $\gamma$, we obtain a wider range of the measurement strength that corresponds to maximum values of $\chi(\rho_{T}^{\textmd{WM}})$. However, the dependence of $\chi(\rho_{T}^{\textmd{WM}})$ on the gravitational interaction strength and measurement strength is non-linear.

Finally, in  Fig.~\ref{figure6} ($b$), we have $\chi(\rho_{T}^{\textmd{WM}})$ in the domain of $(\omega, p)$. Again, we see non-monotonic relationships between the variables and the dense coding capacity. Different ranges of $p$ are optimal when we vary the energy distance $\omega$. Generally, the optimal range of $p$ narrows down as we increase $\omega$. Also, it is worth noticing that Fig. \ref{figure5} has been obtained, assuming a relatively low temperature, which is $T = 0.01$. This fact explains why both plots feature considerably large areas in the parameter space corresponding to optimal values of the dense coding capacity.

\section{Conclusion and outlook}\label{sec6}
In this paper, we studied the quantum dense coding with gravcat states under the effects of temperature, the strength of gravitational interaction coupling, and the energy difference between the ground and the excited states. Based on the obtained results, although the damaging effects of temperature on the quantum advantage of dense coding cannot be ignored, the gravitational interaction and energy difference were able to significantly improve the dense coding capacity. Moreover, we discussed the influence of QWM protocol on the dense coding capacity.  Interestingly, we found that for certain values of measurement strength, the dense coding capacity under the weak measurement is greater than that without weak measurement.

The quantum dense coding with gravcat states could have unique applications in secure communication protocols, involving both quantum information processing and manipulation of gravitational fields. We believe that research like the present work, which can leverage both quantum and gravitational phenomena, may open new opportunities in areas such as quantum sensing, quantum communications, and potentially even quantum computing by considering gravitational effects.

\vspace{20pt}

\section*{Acknowledgments}
This research is supported by the Postdoc grant of the Semnan University under Contract No. 21270.
\\
\\
\section*{Disclosures}
The authors declare that they have no known competing financial interests.
\\
\\
\section*{Data availability}
No datasets were generated or analyzed during the current study.

\end{document}